\documentclass[final,5p,times,sort&compress]{elsarticle}
\usepackage{etoolbox}
\usepackage{lineno}
\usepackage{xspace} 
\usepackage[dvipsnames]{xcolor}
\usepackage{amsmath,amsfonts,amssymb}
\usepackage{mathtools}
\usepackage{bm}
\usepackage{siunitx}
\usepackage{textcomp}
\usepackage{forest}
\usepackage{booktabs}
\usepackage{graphbox}
\usepackage{hyperref}
\hypersetup{
     colorlinks   = true,
     citecolor    = gray
}
\usepackage[capitalize]{cleveref}
\usepackage{doi}
\graphicspath{{./}{./figures/}{../figures/}}
\DeclareGraphicsExtensions{.pdf,.png}

\newcommand{\ie}{\mbox{i.\hspace{1pt}e.}\xspace}

\newcommand{\N}[1]{\ensuremath{N_\text{#1}}}
\newcommand{\Ninput}{\N{input}}
\newcommand{\Nhidden}{\N{hidden}}
\newcommand{\Ndata}{\N{training}}
\newcommand{\Ncorrect}{\N{correct}}
\newcommand{\Nwrong}{\N{wrong}}

\newcommand{\Fscore}{\(F\)-score}

\newcommand{\vctr}[1]{\ensuremath{\bm{#1}}}
\newcommand{\set}[1]{\ensuremath{\bm{#1}}}

\newcommand{\avg}[1]{\ensuremath{\overline{#1}}}
\newcommand{\norm}[2][]{\ensuremath{\left|{#2}\right|_{#1}}}
\newcommand{\pcnt}[1]{\SI{#1}{\percent}}

\newcommand{\LM}{\ensuremath{m^\prime_{\alpha\beta}}}
\newcommand{\residualB}{\ensuremath{\Delta b_{\alpha\beta}}}
\newcommand{\SFsum}{\ensuremath{\text{SF}_\alpha+\text{SF}_\beta}}
\newcommand{\mproduct}{\ensuremath{\LM\,(\SFsum)}}
\newcommand{\mdivide}{\ensuremath{\log(\LM/\residualB)}}

\newcommand{\burgerslength}{\ensuremath{b}}

\DeclareSIUnit\burgersunit{$\burgerslength$}
\DeclareSIUnit\micron{\micro\metre}

\journal{Scripta Materialia}

\begin{document}

\begin{frontmatter}


\title{Grain boundary slip transfer classification and metric selection with artificial neural networks}

\author[MSU]{Zhuowen Zhao}
\author[MSU]{Thomas R. Bieler}
\author[IMDEA,UMadrid]{Javier LLorca}
\author[MSU]{Philip Eisenlohr\corref{PE}}

\ead{eisenlohr@egr.msu.edu}
\cortext[PE]{Corresponding author. Tel: +1 517 4324506}

\address[MSU]{
Chemical Engineering and Materials Science,
Michigan State University,
East Lansing, MI 48824, USA
}

\address[IMDEA]{
IMDEA Materials Institute, 28906 Getafe, Madrid, Spain
}

\address[UMadrid]{
Department of Materials Science, Polytechnic University of Madrid, E. T. S. de Ingenieros de Caminos, Canales y Puertos, 28040 Madrid, Spain
}


\begin{abstract}
	An artificial neural network is used to evaluate the effectiveness of six metrics and their combinations to assess whether slip transfers across grain boundaries in coarse-grained oligocrystalline Al foils \citep{Bieler_etal2019_2,Alizadeh_etal2020}.
	This approach extends the one- or two-dimensional projections formerly applied to analyze slip transfer. The accuracy of this binary classification reaches around \pcnt{87} for the best single metric and around \pcnt{90} when considering two or more metrics simultaneously.
	The results suggest slip transfer mostly depends on the geometric relationship between grains.
	Training a double-layer network having \num{10} nodes per hidden layer with \num{40} measurements is sufficient to render the maximum accuracy. 
\end{abstract}


\begin{keyword}
	grain boundary, slip transfer, classification, metric selection, artificial neural network
\end{keyword}

\end{frontmatter}



The deformation of polycrystalline materials is heterogeneous because the local stress states are constrained by complex evolving boundary conditions associated with the deformation in neighboring grains \citep{Delaire_etal2000, Kacher_etal2014, Fallahi_etal2006, Bieler_etal2014_3}.
Grain boundaries (GBs) are often considered as strong barriers to dislocations, which lead to the formation of dislocation pile-ups \citep{Bayerschen_etal2016}.
However, it is also possible that slip by an incoming dislocation can be transferred into the neighboring grain by passing through the boundary or by absorbing it into the GB and re-emitting a dislocation on a similarly oriented slip system in the receiving grain \citep{Bayerschen_etal2016, Malyar_etal2017}. 
This process could reduce the buildup of local stress, or weaken the GB by accumulation of residual Burgers vector content that could facilitate crack nucleation at GBs \citep{Bieler_etal2009_2, Boehlert_etal2012, Sangid_etal2011_2, Roters_etal2010}. 
Comparisons of crystal plasticity simulations of carefully characterized experiments demonstrate the greatest disagreement in regions near grain boundaries, implying that introducing GB properties into crystal plasticity models is needed to improve the ability to predict the evolution of local stress and strain near boundaries \citep{Gonzalez_etal2014, DiGioacchino+QuintadaFonseca2015, Lim_etal2015, Lim_etal2016_2, Buchheit_etal2015, Guery_etal2016, Hemery_etal2017, Bond+Zikry2017, Plancher_etal2019, Bieler_etal2019_2, Linne_etal2019}.
Nevertheless, reliable metrics that describe slip transfer occurrence at GBs are not well established \citep{Hemery_etal2018, Bieler_etal2019_2, Alizadeh_etal2020}, and they are critical for implementing this phenomenon into standard crystal plasticity simulations of polycrystals \citep{Bieler_etal2019_2, Haouala_etal2020}. 

Some frequently discussed metrics relevant to slip transfer from slip system $\alpha$ to slip system $\beta$ in the neighboring grain include the misorientation between the two neighboring grains (disorientation angle), Luster--Morris parameter (\LM), and the magnitude of the normalized residual Burgers vector (\residualB) \citep{Lee_etal1989, Luster+Morris1995, Seal_etal2012, Bieler_etal2019_2}.
Recent experimental quantification of slip transfer metrics in pure aluminum reported by \citet{Bieler_etal2019_2} and \citet{Alizadeh_etal2020} indicate that slip transfer was observed at low-angle boundaries when \(\LM > 0.97\) or \(\LM > 0.9\) together with \(\residualB < 0.35\). 
\Citet{Bieler_etal2019_2} also suggested there might exist two ``composite'' metrics, namely \LM\ multiplied by the sum \SFsum\ of Schmid factors of the observed slip systems or divided by \residualB.
The effectiveness of these metrics and their relative importance is presented graphically and assessed with simple statistical comparisons.
An assessment of the interrelationship of three or more metrics is difficult when limited to human perception that most easily detects relationships in a (two-dimensional) space spanned by two metrics.

In this paper, an exhaustive evaluation of the aforementioned metrics and their combinations are performed based on the data of \citet{Bieler_etal2019_2} and \citet{Alizadeh_etal2020} using artificial neural network (ANN) as classifier for slip transfer in order to take advantage of its ability to solve hierarchical problems in multidimensional space by mimicking the behavior of the brain \citep{Puri_etal2016, LeCun_etal2015, White1989}.


\begin{figure*}
\centering
\includegraphics{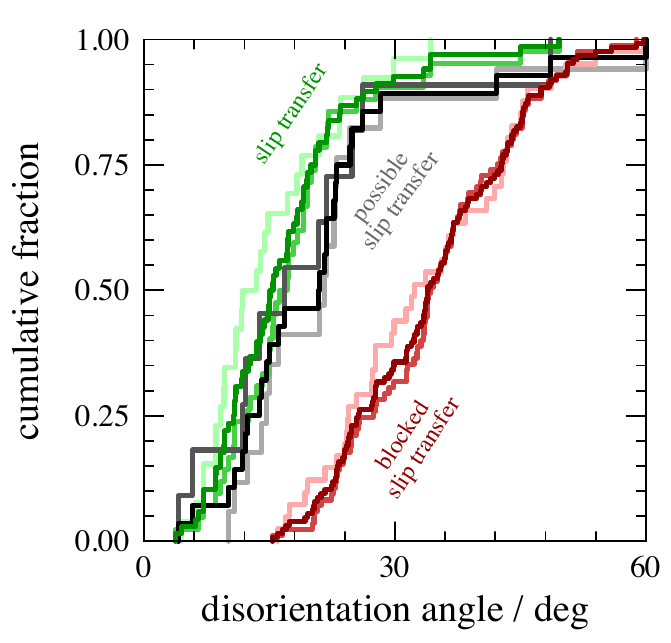}
\includegraphics{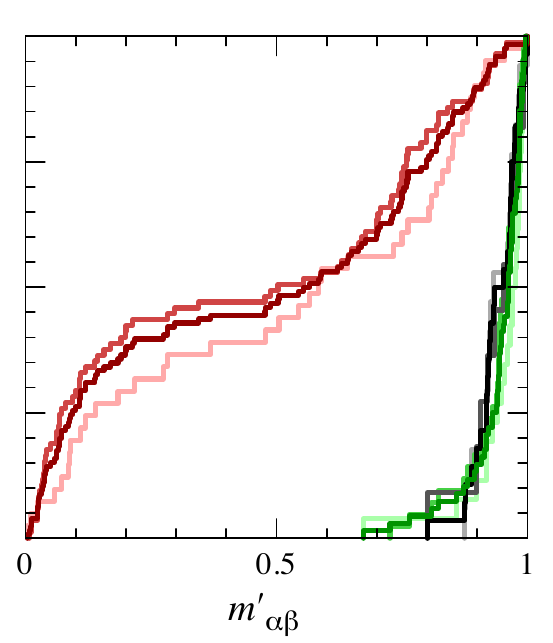}
\includegraphics{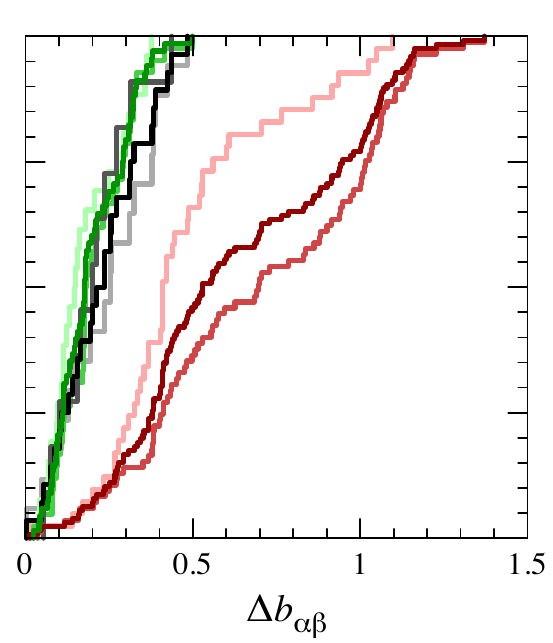}

\includegraphics{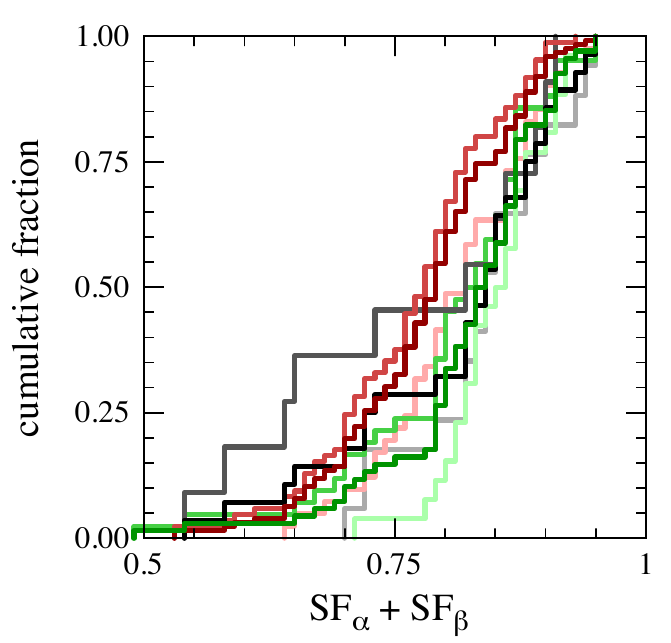}
\includegraphics{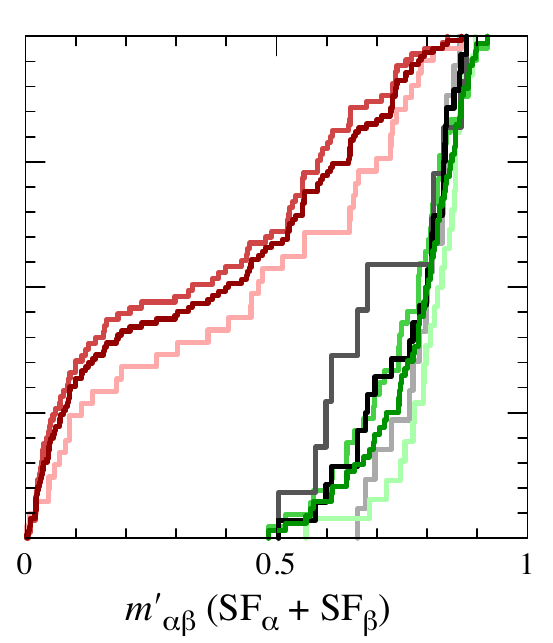}
\includegraphics{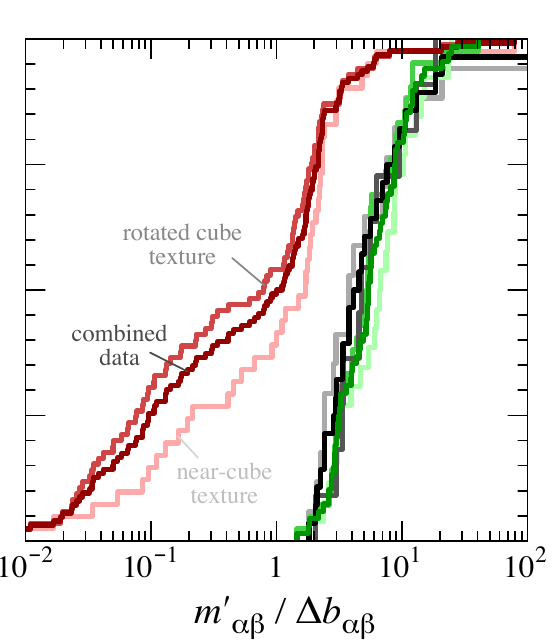}
\caption{
Distributions of metric values classified as either definite, likely, or no slip transfer (green, gray, red).
The underlying population of near-cube texture, rotated-cube texture, and their union are indicated by light, intermediate, and dark color shade, respectively.
} 
\label{fig: distributions}
\end{figure*}

Observations of slip transfer, possible slip transfer, and no slip transfer were collected from 222 GBs across two oligocrystalline Al foils with either a near-cube texture (84 GBs) or a rotated-cube texture (138 GBs) deformed \pcnt{4} in tension at ambient temperature \citep{Bieler_etal2019_2, Alizadeh_etal2020}.
The criteria used to make these observations are discussed at length in \citep{Bieler_etal2019_2} and are briefly noted here:  
Convincing instances of slip transfer occurred when there was highly apparent continuity of slip traces (microshear bands) from one grain to the other grain, with little topography other than slip traces along the grain boundary, implying that the boundary did not cause a significant disruption of slip (the surface between slip bands going across the boundary was smooth).
Instances of no slip transfer continuity of slip across the boundary, but slip bands were still clearly visible in the grain on either side of the boundary; such grains often had significant topographic features along the GB such as a ledge that indicated a discontinuity in strain.  
Cases with less certainty about slip transfer, where evidence of contiguous slip was present together with heterogeneous strain along the boundary, were put in the `possible slip transfer' category, representing about \pcnt{15} of the observations.
The clearest identification of metrics for the activity of slip transfer emerged when non-slip transfer cases were assessed using only the observed slip systems within the neighboring grains (recognizing that some slip systems in neighboring grains had well aligned slip systems, but no  slip activity was observed on these systems).
For each observation, there are four ``direct'' metrics, namely the disorientation angle, Luster--Morris parameter \LM, normalized residual Burgers vector magnitude \(\residualB = 2\norm{\vctr b_\alpha - \vctr b_\beta}/\left(\norm{\vctr b_\alpha}+\norm{\vctr b_\beta}\right)\), and the sum \SFsum\ of Schmid factors of the observed slip systems, and two ``composite'' features, namely \mproduct\ and \mdivide.

\Cref{fig: distributions} compares for each of the six metrics, the cumulative fraction of observations of slip transfer, possible slip transfer, and blocked slip (green, gray, and red, respectively) for both textures individually (two lighter shades) as well as for their combination (darkest shade), resulting in nine populations per metric.
Two observations can be made:
First, the (gray) populations of possible slip transfer are smaller than either of the other sets, and they are consistently close to those of actual slip transfer (green). Second, except for \residualB, both textures result in similar distributions.
Therefore, only the categories of slip transfer and no slip transfer are considered in the subsequent assessment using the combined texture data (\ie, the dark green and dark red distributions, constituting \num{194} out of the \num{222} data).


\begin{figure}
\centering
\includegraphics{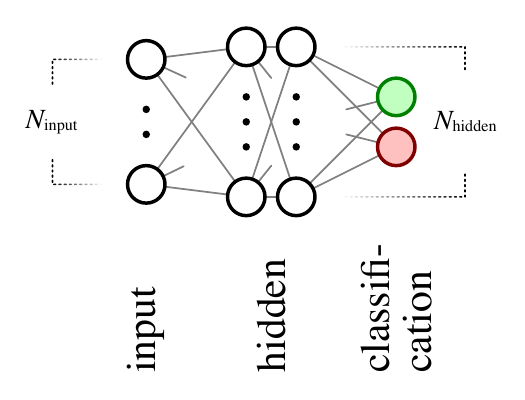}
\caption{Architecture of the artificial neural network used for binary classification (green, red) with \Ninput\ nodes in the input layer and \Nhidden\ nodes in each of the two hidden layers.}
\label{fig: architecture}
\end{figure}

\begin{figure*}
\centering
\setlength\tabcolsep{0 pt}
\begin{tabular}{*{4}{c}@{\hskip 5pt}|@{\hskip 5pt}c}
\(\Ninput = 1\) & 2 & 3 & \(>3\) & variable \Ndata
\\
 \includegraphics[align=t]{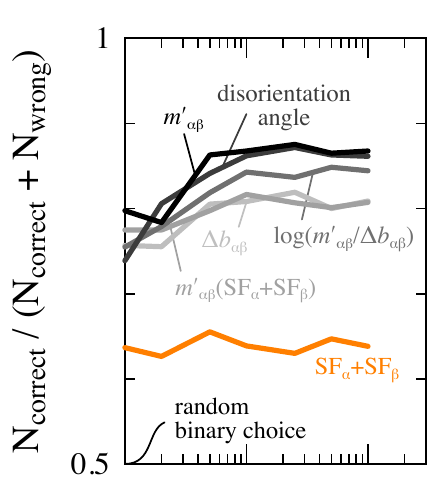}
&\includegraphics[align=t]{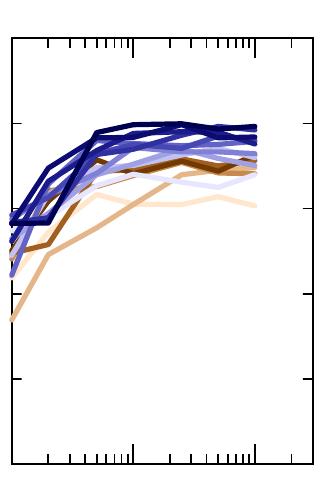}
&\includegraphics[align=t]{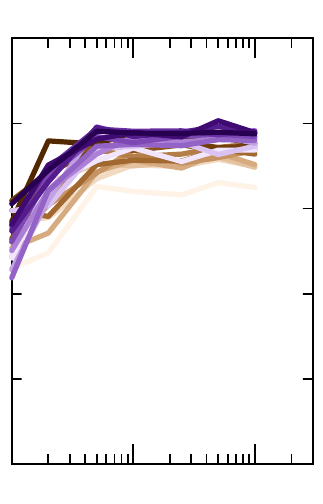}
&\includegraphics[align=t]{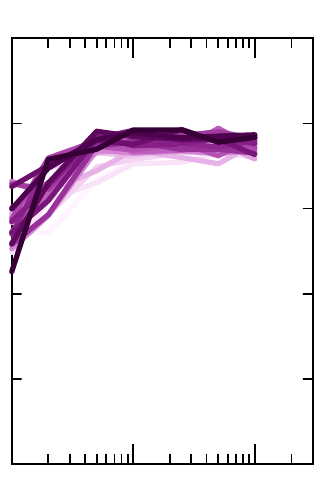}
&\includegraphics[align=t]{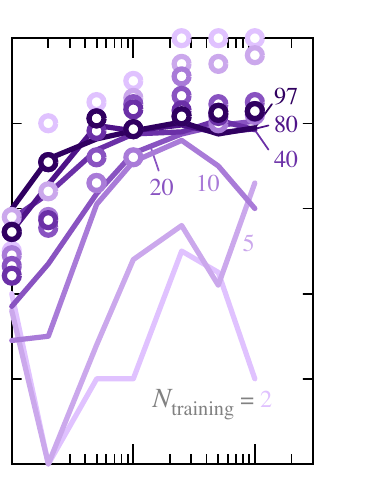}
\\
 \includegraphics[align=b]{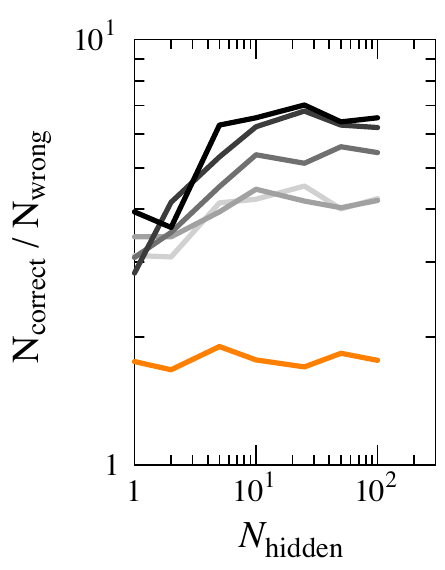}
&\includegraphics[align=b]{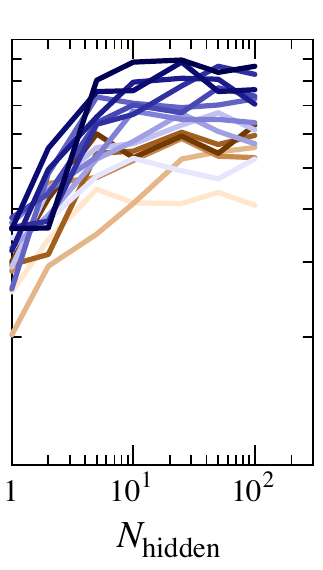}
&\includegraphics[align=b]{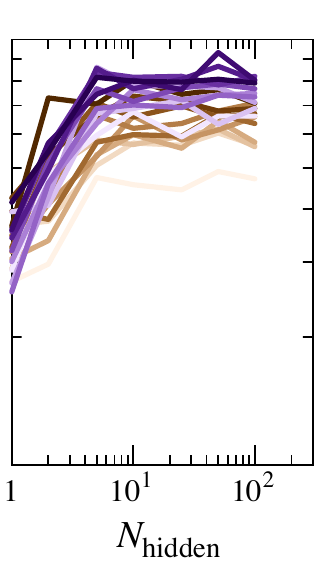}
&\includegraphics[align=b]{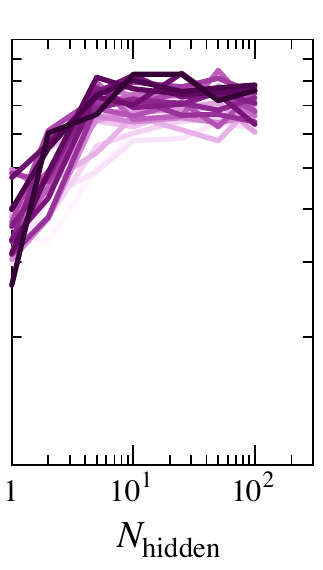}
&\includegraphics[align=b]{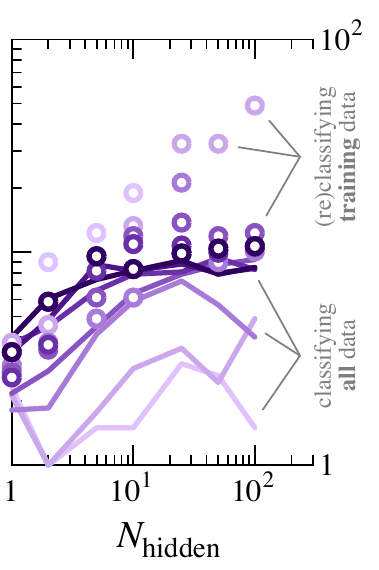}
\\
\end{tabular}
\caption{Classification accuracy as function of hidden layer node count \Nhidden\ for different \Ninput\ of compounded slip transfer metrics with \(\Ndata = 97\).
Both rows present the \emph{same} data but use different ratios (and different scales) to measure accuracy.
Color shading reflects relatively better accuracy from light to dark.
Orange and brownish shading indicates that \SFsum\ are included in the input metric(s).
The right-most column varies \Ndata\ (shades of purple) for one case of \(\Ninput = 3\) and contrasts the classification accuracy of all data (lines) to that of just reclassifying the training data (circles).
} 
\label{fig: accuracies}
\end{figure*}

The biggest problem with applying machine learning on a small training set is a higher risk of overfitting or, equivalently, of poor generalization.
\citet{Feng_etal2019} showed that an ANN with a simple network structure containing only a few hidden layers is potentially effective with small training sets that are quite common for materials science research.
They also demonstrated that a simple ANN does an incrementally better job than a support vector machine (SVM), another commonly used machine learning tool.
This motivated our use of an ANN with two hidden layers (\cref{fig: architecture}).
The appropriateness of using this network is validated in the rightmost plots in \cref{fig: accuracies}---no apparent overfitting is observed in our case when training with more than about 40 data sets.

The complete architecture of the network used in this study is shown in \cref{fig: architecture} and has an input layer with \Ninput\ nodes, two hidden layers containing \Nhidden\ nodes each, and a two-node output layer to classify the input data into either slip transfer being observed or not.
Every edge linearly maps the source node value to the target node.
All edge values are summed and modified by either the (local) rectified linear units (``ReLU'') function within the hidden layer or the (non-local) softmax function at the output layer.
The ultimate classification is based on which of the two output node values is larger.

One training cycle of the ANN consists of forward-feeding all data sets and reducing the value of the loss function (squared \(L_2\) norms of the respective differences between predicted and correct classification) by gradient descent, \ie, updating all edge mapping functions through back-propagation (with a learning rate of \num{e-4}).
Training is terminated when changes to the edge mapping stabilize, which corresponds to roughly \num{2000} cycles (epochs).


The major question addressed in this study is how accurately an ANN can classify slip transfer across GBs based on six metrics related to the crystallography at the boundary and an assumed uniaxial stress state that provides Schmid factor values for each slip system (a measure of the resolved shear stress acting on each slip systems).
Each individual metric and all their various combinations have been tested.
Moreover, the influence of the ANN architecture (\Nhidden) and of the amount of training data (\Ndata) has been examined.
To investigate the architecture, the number of hidden layer nodes was varied as \Nhidden\ = \numlist{1;2;5;10;25;50;100}.
To establish how much training data is necessary, two non-overlapping subsets, each containing \Ndata\ = \numlist{2;5;10;20;40;80;97} data points, are randomly selected from the overall data (\num{194} observations).
The first set is used for training, while the first or the second are utilized to evaluate the classification performance of the trained ANN.
Systematic differences in classification performance obtained using the test data or the training data indicates overfitting (\ie, memorization) of the training data.
\Cref{fig: accuracies} presents the classification accuracy (reflecting the average of 20 independently trained ANN instances) resulting from varying \Ninput, \Nhidden, as well as \Ndata.



Considering the \(\Ninput = 1\) case of \cref{fig: accuracies}, the six individual metrics result in a range of notably different classification accuracy (sorted by shades of gray).
In particular, the disorientation angle and \LM\ exhibit the highest accuracies at \(\Ncorrect/\Nwrong \approx 7\), while the accuracy of \SFsum\ is close to that expected of a random coin toss.

\begin{table}
	\caption{
		\Fscore s of metric value distributions shown in \cref{fig: distributions}.
	}
	\label{tbl: Fscores}
	\centering
	\begin{tabular}{cc} 
	\toprule
	metric & \Fscore \\
	\midrule
	\mproduct & 1.0 \\
	disorientation angle  & 0.97 \\
	\LM & 0.96 \\
	\residualB & 0.84 \\
	\mdivide & 0.75 \\
	\SFsum & 0.063 \\
	\bottomrule
	\end{tabular}
\end{table}
	
These accuracies should reflect the distinctiveness (or separability) of the metric distributions plotted in \cref{fig: distributions}.
The quantity
\begin{align}
	\label{eq: Fscore}
	F &= \frac{\left(\avg{\set{x_{(+)}}}  - \avg{\set{x}}\right)^2 + 
			   \left(\avg{\set{x_{(-)}}}  - \avg{\set{x}}\right)^2}
			   {{\sigma_{(+)}}^2 + {\sigma_{(-)}}^2},
\end{align}
termed \Fscore\ \citep{Chen+Lin2006}, is one method to quantify the ability to discriminate two sets \(\set{x_{(+)}} \cup \set{x_{(-)}} = \set{x}\) of real numbers having unbiased variances of \({\sigma_{(+)}}^2\) and \({\sigma_{(-)}}^2\) and average values of \avg{\set{x_{(+)}}} and \avg{\set{x_{(-)}}}.
Here, \set{x_{(+)}} and \set{x_{(-)}} correspond to observations of slip transfer and blocked slip, \ie, green and red distributions in \cref{fig: distributions}, respectively.
An \Fscore\ of zero is interpreted as \set{x_{(+)}} being inseparable from \set{x_{(-)}} and separability increases with increasing \Fscore.
\Cref{tbl: Fscores} collects the \Fscore s of all six metrics sorted from large (easily separable) to small (not easily separated).
Surprisingly, the relative ranking of the \Fscore s and the ANN ranking are nearly in the opposite order when only a single metric is used for classification (\ie\ \(\Ninput = 1\) in \cref{fig: accuracies} left).
However, the exceptionally low accuracy resulting from the \SFsum\ metric is consistent with its \Fscore\ being almost zero.


Compounding more than one metric as ANN classification inputs results in a slight but notable increase in the observed accuracy compared to the classification based on only a single metric.
The attainable accuracy does not improve after compounding more than two metrics, \ie\ for \(\Ninput > 2\), but the resulting variation decreases with increasing the number of input metrics.
If the metric \SFsum\ is part of the input, the accuracy is systematically worse than without (orange curves in \cref{fig: accuracies}).
This effect diminishes with increasing \Ninput, in line with the overall decrease of the variance.


The attainable classification accuracy saturates at a hidden layer size of 10 nodes and does not change meaningfully up to the maximum investigated \(\Nhidden = 100\).
This indicates that the complexity of the binary slip transfer classification based on one (or more) metrics is low and can be mapped as accurately as possible with an ANN containing a hidden bi-layer of no more than \(2 \times 10\) nodes.


When using subsets with less than \num{40} data sets to train the ANN, the accuracy of the classification resulting from the (unseen) testing data are distinctly lower than those resulting from reclassifying the (already seen) training data.
This discrepancy between lines and circles in \cref{fig: accuracies} (right) is indicative of memorization---in contrast to the desired generalization---by the ANN and becomes more pronounced with lower ratios of \(\Ndata/\Nhidden\).
This is evidenced by (i) the growing deviation between lines and circles with increasing \Nhidden, and (ii) the systematically larger deviation with decreasing \Ndata\ (lighter shades).




Prior work has identified \LM\ and the disorientation angle between grains as metrics that are correlated with observations of slip transfer.
An open question addressed with this study is whether a better slip transfer predictor can be extracted from simultaneously considering multiple metrics using an ANN as classifier, thus avoiding projections to lower dimensions.  

The current ANN-based analysis indicates that no new insights emerge from an objective assessment of interrelationships between multiple metrics. 
Compared to using a single metric, compounding two (or more) metrics as ANN inputs slightly raises the attainable classification accuracy from \pcnt{87} to about \pcnt{90}.
The highest accuracy results from considering metrics that reflect the geometrical relationship between the two grains, \ie\ the disorientation angle and \LM, a widely used metric that encodes geometrical information of slip systems in neighboring grains.
Nevertheless, these two metrics are scalar representations of three-dimensional geometric relationships, so they may have lost information critical to predicting slip transfer more effectively.

Because the accuracy could not be systematically improved by adding more than two input dimensions, these six metrics do not provide sufficient information to enable better prediction.
Nevertheless, increasing the input dimension improves the ``robustness'' of the trained ANN as reflected by the diminishing ``worsening effect'' of the \SFsum\ contribution and the smaller variance of resulting accuracy. 

The low accuracy resulting from the \SFsum\ metric indicates that high global resolved shear stress is insufficient to cause slip transfer.  
This also suggests that use of the global stress state cannot predict the driving forces that actually influence slip transfer at the scale of the grain boundary.  
At that scale, the local stress state is highly variable along the boundary, as shear bands periodically impose large local shear strains that are superimposed by far-field stress states, which evolve in a complex way as influenced by the changing boundary conditions that each grain imposes on its neighbor.
Consequently, knowing the evolution of the local stress state may enable better assessments of driving forces for slip across the length of a particular boundary.  
However, this is difficult to discern, as prior efforts to identify correlations between local stress states and strain evolution near boundaries have not yielded lucid understanding \citep{Gonzalez_etal2014, DiGioacchino+QuintadaFonseca2015, Guery_etal2016, Plancher_etal2019}, suggesting that the history of activated slip systems imposes significant latent hardening effects that could prevent activation of additional slip systems with high resolved shear stresses.  
If so, then commonly used assumptions of plentiful dislocation sources may not be appropriate for predicting the local evolution of strain.
Hence, an incremental approach of installing simple slip transfer mechanisms \citep{Haouala_etal2020, Albiez_etal2019, Bond+Zikry2017, Ma_etal2006_2} with strong latent hardening rules into crystal plasticity models to identify how they alter the evolution of the local stress state may provide insights about causes of the complex evolution of local stresses.
If models of characterized experiments, such as those behind the data used here, are used to identify local stress states, locally informed Schmid factors may identify more convincingly conditions where slip transfer did and did not occur.
With such improvements, it may be possible to reduce the uncertainty in identifying a criterion for slip transfer.

ZZ gratefully acknowledges financial support by the U.S.~Department of Energy through grant DE‐SC0001525.
JLL acknowledges the support from the European Research Council (ERC) under the European Union's Horizon 2020 research and innovation program (Advanced Grant VIRMETAL, grant agreement No.~669141) and from the HexaGB project supported by the Spanish Ministry of Science (reference RTI2018-098245).
This work was supported in part by Michigan State University through computational resources provided by the Institute for Cyber-Enabled Research.

\bibliographystyle{ScriptaMater_doi}
\bibliography{Zhao_etal2020}

\begin{thebibliography}{34}
\providecommand{\natexlab}[1]{#1}
\providecommand{\url}[1]{\texttt{#1}}
\providecommand{\urlprefix}{URL }
\expandafter\ifx\csname urlstyle\endcsname\relax
  \providecommand{\doi}[1]{doi:\discretionary{}{}{}#1}\else
  \providecommand{\doi}{doi:\discretionary{}{}{}\begingroup
  \urlstyle{rm}\Url}\fi

\bibitem[{Bieler et~al.(2019)Bieler, Alizadeh, Pe{\~{n}}a-Ortega, and
  Llorca}]{Bieler_etal2019_2}
T.R. Bieler, R.~Alizadeh, M.~Pe{\~{n}}a-Ortega, J.~Llorca.
\newblock International Journal of Plasticity 118 (2019) 269--290.
\newblock \doi{10.1016/j.ijplas.2019.02.014}.

\bibitem[{Alizadeh et~al.(2020)Alizadeh, Pe{\~{n}}a-Ortega, Bieler, and
  LLorca}]{Alizadeh_etal2020}
R.~Alizadeh, M.~Pe{\~{n}}a-Ortega, T.R. Bieler, J.~LLorca.
\newblock Scripta Materialia 178 (2020) 408--412.
\newblock \doi{10.1016/j.scriptamat.2019.12.010}.

\bibitem[{Delaire et~al.(2000)Delaire, Raphanel, and Rey}]{Delaire_etal2000}
F.~Delaire, J.L. Raphanel, C.~Rey.
\newblock Acta Materialia 48~(5) (2000) 1075--1087.
\newblock \doi{10.1016/S1359-6454(99)00408-5}.

\bibitem[{Kacher et~al.(2014)Kacher, Eftink, Cui, and
  Robertson}]{Kacher_etal2014}
J.~Kacher, B.P. Eftink, B.~Cui, I.M. Robertson.
\newblock Current Opinion in Solid State and Materials Science 18~(4) (2014)
  227--243.
\newblock \doi{10.1016/j.cossms.2014.05.004}.

\bibitem[{Fallahi et~al.(2006)Fallahi, Mason, Kumar, Bieler, and
  Crimp}]{Fallahi_etal2006}
A.~Fallahi, D.E. Mason, D.~Kumar, T.R. Bieler, M.A. Crimp.
\newblock Materials Science and Engineering A 432~(1-2) (2006) 281--291.
\newblock \doi{doi:10.1016/j.msea.2006.06.046}.

\bibitem[{Bieler et~al.(2014)Bieler, Eisenlohr, Zhang, Phukan, and
  Crimp}]{Bieler_etal2014_3}
T.R. Bieler, P.~Eisenlohr, C.~Zhang, H.~Phukan, M.A. Crimp.
\newblock Current Opinion in Solid State and Materials Science 18 (2014)
  212--226.
\newblock \doi{10.1016/j.cossms.2014.05.003}.

\bibitem[{Bayerschen et~al.(2016)Bayerschen, McBride, Reddy, and
  B{\"o}hlke}]{Bayerschen_etal2016}
E.~Bayerschen, A.T. McBride, B.D. Reddy, T.~B{\"o}hlke.
\newblock Journal of Materials Science 51~(5) (2016) 2243--2258.
\newblock \doi{10.1007/s10853-015-9553-4}.

\bibitem[{Malyar et~al.(2017)Malyar, Micha, Dehm, and
  Kirchlechner}]{Malyar_etal2017}
N.V. Malyar, J.S. Micha, G.~Dehm, C.~Kirchlechner.
\newblock Acta Materialia 129 (2017) 312--320.
\newblock \doi{10.1016/j.actamat.2017.03.003}.

\bibitem[{Bieler et~al.(2009)Bieler, Eisenlohr, Roters, Kumar, Mason, Crimp,
  and Raabe}]{Bieler_etal2009_2}
T.R. Bieler, P.~Eisenlohr, F.~Roters, D.~Kumar, D.E. Mason, M.A. Crimp,
  D.~Raabe.
\newblock International Journal of Plasticity 25~(9) (2009) 1655--1683.
\newblock \doi{10.1016/j.ijplas.2008.09.002}.

\bibitem[{Boehlert et~al.(2012)Boehlert, Chen, Guti{\'e}rrez-Urrutia, Llorca,
  and P{\'e}rez-Prado}]{Boehlert_etal2012}
C.J. Boehlert, Z.~Chen, I.~Guti{\'e}rrez-Urrutia, J.~Llorca, M.T.
  P{\'e}rez-Prado.
\newblock Acta Materialia 60~(4) (2012) 1889--1904.
\newblock \doi{10.1016/j.actamat.2011.10.025}.

\bibitem[{Sangid et~al.(2011)Sangid, Maier, and Sehitoglu}]{Sangid_etal2011_2}
M.D. Sangid, H.J. Maier, H.~Sehitoglu.
\newblock International Journal of Plasticity 27~(5) (2011) 801--821.
\newblock \doi{10.1016/j.ijplas.2010.09.009}.

\bibitem[{Roters et~al.(2010)Roters, Eisenlohr, Hantcherli, Tjahjanto, Bieler,
  and Raabe}]{Roters_etal2010}
F.~Roters, P.~Eisenlohr, L.~Hantcherli, D.D. Tjahjanto, T.R. Bieler, D.~Raabe.
\newblock Acta Materialia 58 (2010) 1152--1211.
\newblock \doi{10.1016/j.actamat.2009.10.058}.

\bibitem[{Gonzalez et~al.(2014)Gonzalez, Simonovski, Withers, and Quinta~da
  Fonseca}]{Gonzalez_etal2014}
D.~Gonzalez, I.~Simonovski, P.J. Withers, J.~Quinta~da Fonseca.
\newblock International Journal of Plasticity 61 (2014) 49--63.
\newblock \doi{10.1016/j.ijplas.2014.03.012}.

\bibitem[{Di~Gioacchino and Quinta~da
  Fonseca(2015)}]{DiGioacchino+QuintadaFonseca2015}
F.~Di~Gioacchino, J.~Quinta~da Fonseca.
\newblock International Journal of Plasticity 74 (2015) 92--109.
\newblock \doi{10.1016/j.ijplas.2015.05.012}.

\bibitem[{Lim et~al.(2015)Lim, Hale, Zimmerman, Battaile, and
  Weinberger}]{Lim_etal2015}
H.~Lim, L.M. Hale, J.A. Zimmerman, C.C. Battaile, C.R. Weinberger.
\newblock International Journal of Plasticity 73 (2015) 100--118.
\newblock \doi{10.1016/j.ijplas.2014.12.005}.

\bibitem[{Lim et~al.(2016)Lim, Dingreville, Deibler, Buchheit, and
  Battaile}]{Lim_etal2016_2}
H.~Lim, R.~Dingreville, L.A. Deibler, T.E. Buchheit, C.C. Battaile.
\newblock Computational Materials Science 117 (2016) 437--444.
\newblock \doi{10.1016/j.commatsci.2016.02.022}.

\bibitem[{Buchheit et~al.(2015)Buchheit, Carroll, Clark, and
  Boyce}]{Buchheit_etal2015}
T.E. Buchheit, J.D. Carroll, B.G. Clark, B.L. Boyce.
\newblock Microscopy and Microanalysis 21~(4) (2015) 969--984.
\newblock \doi{10.1017/S1431927615000677}.

\bibitem[{Guery et~al.(2016)Guery, Hild, Latourte, and Roux}]{Guery_etal2016}
A.~Guery, F.~Hild, F.~Latourte, S.~Roux.
\newblock International Journal of Plasticity
  \doi{10.1016/j.ijplas.2016.01.008}.

\bibitem[{H{\'e}mery et~al.(2017)H{\'e}mery, Nait-Ali, and
  Villechaise}]{Hemery_etal2017}
S.~H{\'e}mery, A.~Nait-Ali, P.~Villechaise.
\newblock Mechanics of Materials 109 (2017) 1--10.
\newblock \doi{10.1016/j.mechmat.2017.03.013}.

\bibitem[{Bond and Zikry(2017)}]{Bond+Zikry2017}
D.M. Bond, M.A. Zikry.
\newblock Journal of Engineering Materials and Technology 139~(2).
\newblock \doi{10.1115/1.4035494}.

\bibitem[{Plancher et~al.(2019)Plancher, Tajdary, Auger, Castelnau, Favier,
  Loisnard, Marijon, Maurice, Michel, Robach, and Stodolna}]{Plancher_etal2019}
E.~Plancher, P.~Tajdary, T.~Auger, O.~Castelnau, V.~Favier, D.~Loisnard, J.B.
  Marijon, C.~Maurice, V.~Michel, O.~Robach, J.~Stodolna.
\newblock JOM 71~(10) (2019) 3543--3551.
\newblock \doi{10.1007/s11837-019-03711-5}.

\bibitem[{Linne et~al.(2019)Linne, Venkataraman, Sangid, and
  Daly}]{Linne_etal2019}
M.A. Linne, A.~Venkataraman, M.D. Sangid, S.~Daly.
\newblock Experimental Mechanics 59~(5) (2019) 643--658.
\newblock \doi{10.1007/s11340-019-00517-z}.

\bibitem[{H{\'e}mery et~al.(2018)H{\'e}mery, Nizou, and
  Villechaise}]{Hemery_etal2018}
S.~H{\'e}mery, P.~Nizou, P.~Villechaise.
\newblock Materials Science and Engineering: A 709 (2018) 277--284.
\newblock \doi{10.1016/j.msea.2017.10.058}.

\bibitem[{Haouala et~al.(2020)Haouala, Alizadeh, Bieler, Segurado, and
  LLorca}]{Haouala_etal2020}
S.~Haouala, R.~Alizadeh, T.R. Bieler, J.~Segurado, J.~LLorca.
\newblock International Journal of Plasticity 126 (2020) 102600--.
\newblock \doi{10.1016/j.ijplas.2019.09.006}.

\bibitem[{Lee et~al.(1989)Lee, Robertson, and Birnbaum}]{Lee_etal1989}
T.C. Lee, I.M. Robertson, H.K. Birnbaum.
\newblock Scripta Metallurgica 23~(5) (1989) 799--803.
\newblock \doi{10.1016/0036-9748(89)90534-6}.

\bibitem[{Luster and Morris(1995)}]{Luster+Morris1995}
J.~Luster, J.M. Morris.
\newblock Metallurgical and Materials Transactions A 26~(7) (1995) 1745--1756.
\newblock \doi{10.1007/BF02670762}.

\bibitem[{Seal et~al.(2012)Seal, Crimp, Bieler, and Boehlert}]{Seal_etal2012}
J.R. Seal, M.A. Crimp, T.R. Bieler, C.J. Boehlert.
\newblock Materials Science and Engineering A 552 (2012) 61--68.
\newblock \doi{10.1016/j.msea.2012.04.114}.

\bibitem[{Puri et~al.(2016)Puri, Solanki, Padawer, Tipparaju, Moreno, and
  Pathak}]{Puri_etal2016}
M.~Puri, A.~Solanki, T.~Padawer, S.M. Tipparaju, W.A. Moreno, Y.~Pathak.
\newblock {Introduction to Artificial Neural Network (ANN) as a Predictive Tool
  for Drug Design, Discovery, Delivery, and Disposition}.
\newblock Elsevier (2016) 3--13.
\newblock \doi{10.1016/B978-0-12-801559-9.00001-6}.

\bibitem[{LeCun et~al.(2015)LeCun, Bengio, and Hinton}]{LeCun_etal2015}
Y.~LeCun, Y.~Bengio, G.~Hinton.
\newblock Nature 521~(7553) (2015) 436--444.
\newblock \doi{10.1038/nature14539}.

\bibitem[{White(1989)}]{White1989}
H.~White.
\newblock Neural Computation 1~(4) (1989) 425--464.
\newblock \doi{10.1162/neco.1989.1.4.425}.

\bibitem[{Feng et~al.(2019)Feng, Zhou, and Dong}]{Feng_etal2019}
S.~Feng, H.~Zhou, H.~Dong.
\newblock Materials {\&} Design 162 (2019) 300--310.
\newblock \doi{10.1016/j.matdes.2018.11.060}.

\bibitem[{Chen and Lin(2006)}]{Chen+Lin2006}
Y.W. Chen, C.J. Lin.
\newblock {Combining SVMs with Various Feature Selection Strategies}, vol. 207.
\newblock Springer Berlin Heidelberg, Berlin, Heidelberg (2006) 315--324.
\newblock \doi{10.1007/978-3-540-35488-8_13}.

\bibitem[{Albiez et~al.(2019)Albiez, Erdle, Weygand, and
  B{\"o}hlke}]{Albiez_etal2019}
J.~Albiez, H.~Erdle, D.~Weygand, T.~B{\"o}hlke.
\newblock International Journal of Plasticity 113 (2019) 291--311.
\newblock \doi{10.1016/j.ijplas.2018.10.006}.

\bibitem[{Ma et~al.(2006)Ma, Roters, and Raabe}]{Ma_etal2006_2}
A.~Ma, F.~Roters, D.~Raabe.
\newblock Acta Materialia 54~(8) (2006) 2181--2194.
\newblock \doi{10.1016/j.actamat.2006.01.004}.

\end{thebibliography}

\end{document}